\documentclass[12pt]{iopart_modified}
\usepackage{setspace}

\newcommand{\enlarge}{}

\usepackage{graphicx}

\makeatletter
\renewcommand{\@cite}[2]{({#1\if@tempswa , #2\fi})}
\renewcommand{\@biblabel}[1]{}
\makeatother

\newcommand{\ea}{{\emph{et al.}~}}

\newcommand{\sigmaS}{\sigma_s}
\newcommand{\drho}{\delta\rho}
\newcommand{\eps} {\varepsilon}
\newcommand{\la}{\lambda}
\newcommand{\La}{{\bf\Lambda}}
\newcommand{\Labar}{\bar{\bf\Lambda}}
\newcommand{\dLa}{\delta\La}
\newcommand{\weyl}{{\mbox{\tiny Weyl}}}
\newcommand{\cO}{{\cal O}}

\newcommand{\cN}{{\cal N}}
\newcommand{\cV}{{\cal V}}

\newcommand{\cS}{{\cal S}}
\newcommand{\cC}{{\cal C}}
\newcommand{\cJ}{{\cal J}}

\newcommand{\cT}{{\cal T}}

\newcommand{\unitMat}{\hbox{${\bf 1}$}\raise 0.4pt\hbox{$\hskip -1.9pt\vrule depth 0pt height 1.55ex width 0.7pt\vrule depth 0.0pt height 0.3pt width 0.15em$}}
\newcommand{\del}[1]{\frac{\partial}{\partial #1}}
\newcommand{\Del}[2]{\frac{\partial #1}{\partial #2}}
\newcommand{\ddel}[1]{\frac{\partial^2}{\partial {#1}^2}}
\newcommand{\Rnull}{R_{\mbox{\tiny0}}}
\newcommand{\Jmu}{\cJ_{\!\mu}}

\newcommand{\onlinecite}[1]{\cite{#1}}
\newcommand{\text}[1]{{\rm #1}}
\newcommand{\binom}[2]{\left(\begin{array}{cc} #1 \\ #2 \end{array}\right)}

\begin{document}

\title{The Nanoscale Free-Electron Model}


\author{D.~F.~Urban$^{1,\dagger}$, J.~B\"urki$^{2}$,
C.~A.~Stafford$^3$, and Hermann~Grabert$^{1,4}$}

\address{$^{1}$Physikalisches Institut, Albert-Ludwigs-Universit\"at, Hermann-Herder-Stra{\ss}e 3, D-79104 Freiburg, Germany}
\address{$^{2}$Department of Physics and Astronomy, California State University, 6000 J Street, Sacramento,
\mbox{CA 95819-6041}, USA}
\address{$^{3}$Department of Physics, University of Arizona, 1118 E. Fourth Street, Tucson, \mbox{AZ 85721}, USA}
\address{$^{4}$Freiburg Institute for Advanced Studies, Albert-Ludwigs-Universit\"at, Albertstra{\ss}e 19, D-79104 Freiburg}

\address{$^{\dagger}$Email: urban@physik.uni-freiburg.de}

\begin{abstract}
A brief review of the nanoscale free-electron model of metal
nanowires is presented. This continuum description of metal
nanostructures allows for a unified treatment of cohesive and
conducting properties. Conductance channels act as delocalized
chemical bonds whose breaking is responsible for jumps in the
conductance and force oscillations. It is argued that surface and
quantum-size effects are the two dominant factors in the
energetics of a nanowire, and much of the phenomenology of
nanowire stability and structural dynamics can be understood
based on the interplay of these two competing factors. A linear
stability analysis reveals a sequence of ``magic'' conductance
values for which the underlying nanowire geometry is
exceptionally stable. The stable configurations include
Jahn-Teller deformed wires of broken axial symmetry. The model
naturally explains the experimentally observed shell and
supershell structures.
\end{abstract}

\maketitle

\setcounter{footnote}{0}
\newpage
\setcounter{page}{1} \tableofcontents
\newpage

\section{Introduction}\label{sec:intro}

The past decades have seen an accelerating miniaturization of both
mechanical and electrical devices, so that a better understanding
of properties of ultrasmall systems is required in increasing
detail. The first measurements of conductance quantization in the
late 1980s \cite{Wees88,Wharam88} in constrictions of
two-dimensional electron gases formed by means of gates, have
demonstrated the importance of quantum confinement effects in
these systems and opened a wide field of research. A major step
has been the discovery of conductance quantization in metallic
nanocontacts \cite{Agrait93,Brandbyge95,Krans95}: The conductance
measured during the elongation of a metal nanowire is a steplike
function where the typical step height is frequently near a
multiple of the conductance quantum $G_0=2e^2/h$, where $e$ is
the electron charge and $h$  Planck's constant. Surprisingly,
this was initially not interpreted as a quantum effect but rather
as a consequence of abrupt atomic rearrangements and elastic
deformation stages. This interpretation, supported by a series of
molecular dynamics simulations \cite{Landman90,Todorov93}, was
claimed to be confirmed by another pioneering experiment
\cite{Rubio96,Stalder96} measuring simultaneously the conductance
and the cohesive force of gold nanowires with diameters ranging
from several {\AA}ngstroms to several nanometers.  As the contact
was pulled apart, oscillations in the force of order 1nN were
observed in perfect correlation with the conductance steps.

It came as a surprise when Stafford \ea(1997) introduced the
free-electron model of a nanocontact -- referred to as the
\emph{Nanoscale Free-Electron Model} (NFEM) henceforth -- and
showed that this comparatively simple model, which emphasizes the
quantum confinement effects of the metallic electrons, is able to
reproduce quantitatively the main features of the experimental
observations. In this approach, the nanowire is understood to act
as a quantum waveguide for the conduction electrons (which are
responsible for both conduction and cohesion in simple metals):
Each quantized mode transmitted through the contact contributes
$G_0$ to the conductance and a force of order $E_F/\la_F$ to the
cohesion, where $E_F$ and $\la_F$ are the Fermi energy and
wavelength, respectively. Conductance channels act as delocalized
bonds whose stretching and breaking is responsible for the
observed force oscillations, thus explaining straightforwardly
their correlations with the conductance steps.

Since then, free-standing metal nanowires, suspended from
electrical contacts at their ends, have been fabricated by a
number of different techniques. Metal wires down to a single atom
thick were extruded using a scanning-tunneling microscope tip
\cite{Rubio96,Untiedt97}. Metal nanobridges were shown to
``self-assemble'' under electron-beam irradiation of thin metal
films (Kondo \ea 1997, 2000, Rodrigues \ea 2000), leading to
nearly perfect cylinders down to four atoms in diameter, with
lengths up to fifteen nanometers. In particular, the
mechanically-controllable break junction technique, introduced by
Moreland and Ekin (1985) and refined by Ruitenbeek and coworkers
\cite{Muller92}, has allowed for systematic studies of nanowire
properties for a variety of materials. For a survey see the
review by Agra{\"\i}t \ea (2003).

A remarkable feature of metal nanowires is that they are stable
at all. Most atoms in such a thin wire are at the surface, with
small coordination numbers, so that {\em surface effects} play a
key role in their energetics. Indeed, macroscopic arguments
comparing the surface-induced stress to the yield strength
indicate a minimum radius for solidity of order ten nanometers
\cite{Zhang03}. Below this critical radius and absent some other
stabilizing mechanism, plastic flow would lead to a Rayleigh
instability \cite{Chandrasekhar81} breaking the wire apart into
clusters. Already in the 19th century Plateau (1873) realized
that this surface-tension-driven instability is unavoidable if
cohesion is due solely to classical pairwise interactions between
atoms. The experimental evidence accumulated over the past decade
on the remarkable stability of nanowires considerably thinner
than the above estimate clearly shows that electronic effects
emphasized by the NFEM dominate over atomistic effects for
sufficiently small radii.

A series of experiments on alkali metal nanocontacts (Yanson \ea
1999, 2001) identified {\em electron-shell effects}, which
represent the semiclassical limit of the quantum-size effects
discussed above, as a key mechanism influencing nanowire
stability. Energetically-favorable structures were revealed as
peaks in conductance histograms, periodic in the nanowire radius,
analogous to the electron-shell structure previously observed in
metal clusters \cite{deHeer93}. A supershell structure was also
observed \cite{Yanson00}, in the form of a periodic modulation of
the peak heights. More recently, such electron-shell effects have
also been observed, even at room temperature, for the noble metals
gold, copper, and silver (Diaz \ea 2003, Mares \ea 2004, 2005) as
well as for aluminum \cite{Mares07}.

Soon after the first experimental evidence for electron shell
effects in metal nanowires, a theoretical analysis using the NFEM
found that nanowire stability can be explained by a competition
of the two key factors, surface tension and electron-shell
effects \cite{Kassubek01}. Both linear \cite{Zhang03,Urban03} and
nonlinear (B\"urki \ea 2003, 2005a) stability analyses of axially
symmetric nanowires found that the surface-tension driven
instability can be completely suppressed in the vicinity of
certain ``magic radii.'' However, the restriction to axial
symmetry implies characteristic gaps in the sequence of stable
nanowires, which is not fully consistent with the experimentally
observed nearly perfect periodicity of the conductance peak
positions. A Jahn-Teller deformation breaking the symmetry can
lead to more stable deformed configurations. Recently, the linear
stability analysis was extended to wires with arbitrary
cross-section (Urban \ea 2004a, 2006). This general analysis
confirms the existence of  a sequence of magic cylindrical wires
of exceptional stability which represent roughly 75\% of the main
structures observed in conductance histograms. The remaining 25\%
are deformed and predominantly of elliptical or quadrupolar
shapes. This result allows for a consistent interpretation of
experimental conductance histograms for alkali and noble metals,
including both the electronic shell and supershell structures
\cite{Urban04b}.

This chapter is intended to give an introduction to the NFEM.
Section \ref{sec:model} summarizes the assumptions and features of
the model while the general formalism is described in Sec.\
\ref{sec:formalism}. In the following sections, two applications
of the NFEM will be discussed: First, we give a unified
explanation of electrical transport and cohesion in metal
nanocontacts (Sec.\ \ref{sec:ForceCond}) and second, the linear
stability analysis for straight metal nanowires will be presented
(Sec.\ \ref{sec:StabAna}). The latter will include cylindrical
wires as well as wires with broken axial symmetry, thereby
discussing the Jahn-Teller-effect.

\section{Assumptions and limitations of the NFEM}\label{sec:model}

Guided by the importance of conduction electrons in the cohesion
of metals, and by the success of the jellium model in describing
metal clusters \cite{deHeer93,Brack93}, the NFEM replaces the
metal ions by a uniform, positively charged background that
provides a confining potential for the electrons. The electron
motion is free along the wire, and confined in the transverse
directions. Usually an infinite confinement potential (hard-wall
boundary conditions) for the electrons is chosen. This is
motivated by the fact that the effective potential confining the
electrons to the wire will be short ranged due to the strong
screening in good metals.

In a first approximation electron-electron interactions are
neglected, which is reasonable due to the excellent screening
\cite{Kassubek99} in metal wires with $G>G_0$. It is known from
cluster physics that a free electron model gives qualitative
agreement and certainly describes the essential physics involved.
Interaction, exchange and correlation effects as well as a
realistic confinement potential have to be taken into account,
however, for quantitative agreement.%
\footnote{Note however, that the error introduced by using
hard-walls instead of a more realistic soft-wall confining
potential can be essentially corrected for by placing the
hard-wall a finite distance outside the wire surface, thus
compensating for the over-confinement \cite{Garcia-martin96}.}
From this we infer that the same is true for metal nanowires,
where similar confinement effects are important. Remarkably, the
electron-shell effects crucial to the stabilization of long wires
are described with quantitative accuracy by the simple
free-electron model, as discussed below.

In addition, the NFEM assumes that the positive background behaves
like an incompressible fluid when deforming the nanowire. This
takes into account, to lowest order, the hard-core repulsion of
core electrons as well as the exchange energy of conduction
electrons. When using a hard-wall confinement, the Fermi energy
$E_F$ (or equivalently the Fermi wavelength $\la_F$) is the only
parameter entering the NFEM. As $E_F$ is material dependent and
experimentally accessible, there is no adjustable parameter. This
pleasant feature needs to be abandoned in order to model different
materials more realistically. Different kinds of appropriate
surface boundary conditions are imaginable in order to model the
behaviour of an incompressible fluid and to fit the surface
properties of various metals. This will be discussed in detail in
Sec. \ref{sec:generalconstraint}.

A more refined model of a nanocontact would consider effects of
scattering from disorder (B\"urki \ea 1999a, b) and
electron-electron interaction via a Hartree approximation
\cite{Stafford00,Zhang05}. The inclusion of disorder in particular
leads to a better quantitative agreement with transport
measurements, but does not change the cohesive properties
qualitatively in any significant way, while electron-electron
interactions are found to be a small correction in most cases. As
a result, efforts to make the NFEM more realistic do not improve
it significantly, while removing one of its main strengths, the
absence of any adjustable parameters.

The major shortcoming of the NFEM is that its applicability is
limited to good metals having a nearly spherical Fermi surface. It
is best suited for the (highly reactive) s-orbital alkali metals,
providing a theoretical understanding of the important physics in
nanowires. The NFEM has also been proven to qualitatively (and
often semi-quantitatively) describe noble metal nanowires, and in
particular, gold. Lately, it has been shown that the NFEM can even
be applied (within a certain parameter range) to describe the
multivalent metal aluminum, since Al shows an almost spherical
Fermi surface in the extended-zone scheme.
The NFEM is especially suitable to describe shell effects due to
the conduction-band s-electrons, and the experimental observation
of a crossover from atomic-shell to electron-shell effects with
decreasing radius in both metal clusters \cite{Martin96} and
nanowires \cite{Yanson01} justifies {\it a posteriori} the use of
the NFEM in the later regime.
Naturally, the NFEM does not capture effects originating from the
directionality of bonding, such as the effect of surface
reconstruction observed for Au. For this reason it cannot be used
to model atomic chains of Au atoms, which are currently
extensively
studied experimentally.
Keeping these limitations in mind, the NFEM is applicable within a
certain range of radius, capturing nanowires with only very few
atoms in cross-section up to wires of several nanometers in
thickness, depending on the material under consideration.

\section{Formalism of the NFEM}\label{sec:formalism}

\subsection{Scattering matrix formalism}
\label{sec:SMatrix}

A metal nanowire represents an open system connected to metallic
electrodes at each end.
These macroscopic electrodes 
act as ideal electron reservoirs in thermal equilibrium with a
well-defined temperature and chemical potential. When treating an
open system, the Schr\"odinger equation is most naturally
formulated as a scattering problem. The basic idea of the
scattering approach is to relate physical properties of the wire
with transmission and reflection amplitudes for electrons being
injected from the leads.\footnote{Phase-coherence is assumed to be
preserved in the wire (a good approximation given the size of the
system compared to the inelastic mean-free-path) and inelastic
scattering is restricted to the electron reservoirs only.}

The fundamental quantity describing the properties of the system
is the energy-dependent unitary scattering matrix $S(E)$
connecting incoming and outgoing asymptotic states of conduction
electrons in the electrodes. For a quantum wire, 
$S(E)$ can be decomposed into four submatrices
$S_{\alpha\beta}(E)$, $\alpha$, $\beta=1,2$, where 1 (2)
indicates the left (right) lead.  Each submatrix $S_{\alpha
\beta}(E)$ determines how an incoming eigenmode of lead $\beta$
is scattered into a linear combination of outgoing eigenmodes of
lead $\alpha$. The eigenmodes of the leads are also referred to as
scattering channels.

The formulation of electrical transport in terms of the
scattering matrix was developed by Landauer and B{\"u}ttiker: The
(linear response) electrical conductance $G$ can be expressed as
a function of the submatrix $S_{21}$ which describes transmission
from the source electrode 1 to the drain electrode 2 and is given
by \cite{Datta95}
\begin{equation}
    \label{condformula}
    G = \frac{2e^2}{h} \int dE\, \frac{-\partial f(E)}{\partial E} \mbox{Tr}_1\left\{
    S_{21}^{\dagger}(E) S_{21}(E)\right\}.
\end{equation}
Here $f(E)=\{\exp[\beta(E-\mu)]+1\}^{-1}$ is the Fermi
distribution function for electrons in the reservoirs,
$\beta=(k_BT)^{-1}$ is the inverse temperature and $\mu$ is the
electron chemical potential, specified by the macroscopic
electrodes. The trace $\mbox{Tr}_1$ sums over all eigenmodes of
the source.

The appropriate thermodynamic potential to describe the
energetics of an open system is the grand canonical potential
\begin{equation}
\label{eq:OmegaVonD}
    \Omega=-\frac{1}{\beta} \int \!dE\, D(E) \,
    \ln\!\left[1+e^{-\beta(E-\mu)}
        \right],
\end{equation}
where $D(E)$ is the electronic density of states (DOS) of the
nanowire. Notably, the DOS of an open system may also be expressed
in terms of the scattering matrix as \cite{Dashen69}
\begin{eqnarray}
\label{eq:DausS}
        D(E) &=&
    \frac{1}{2\pi i}\;\mbox{Tr}\left\{
    S^{\dagger}(E)\frac{\partial S}{\partial E} -
        \frac{\partial S^{\dagger}}{\partial E}S(E)\right\},
\end{eqnarray}
where Tr sums over the states of both electrodes. This formula is
also known as Wigner delay. Note that Eqs.~(\ref{condformula}),
(\ref{eq:OmegaVonD}), and (\ref{eq:DausS}) include a factor of 2
for spin degeneracy.

Thus, once the electronic scattering problem for the nanowire is
solved, both transport and energetic quantities can be readily
calculated.

\subsection{WKB approximation}
\label{sec:WKB}

\begin{figure}[t]
\begin{center}
    \includegraphics[width=0.75\columnwidth,draft=false]{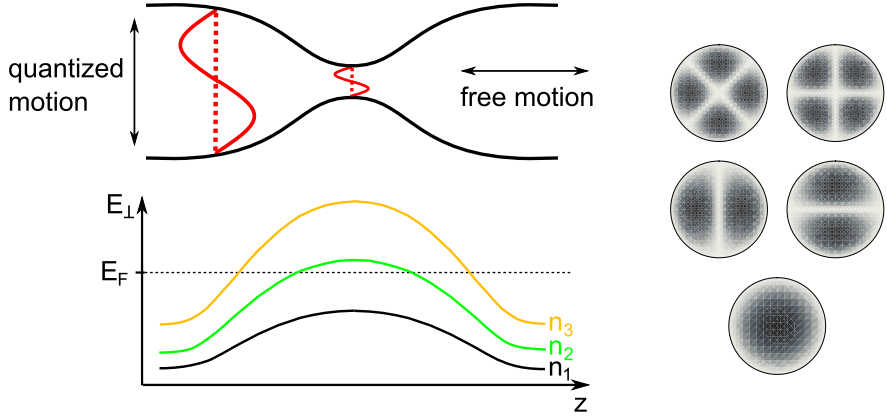}
    \caption[]{
     Upper-left part: Sketch of a nano-constriction. Within the adiabatic
     approximation, transverse and longitudinal motions are separable:
     the motion in the transverse direction is quantized, while in the
     longitudinal direction the electrons move in a potential created
     by the transverse energies [see Eq.\ (\ref{eq:wkb:schroedingerC})].
     Lower-left part: Sketch of transverse
     energies for different transverse channels $n_1$, $n_2$, and $n_3$ as a
     function of the $z$-coordinate. Channel $n_1$ is transmitted
     through the constriction as its maximum transverse energy is
     smaller than the Fermi energy, channel $n_2$ is partly transmitted,
     and channel $n_3$ is almost totally reflected.
     Right part: Density plots of $|\Psi_n(r,\phi)|^2$ for the three eigenmodes
     depicted on the lower-left part, corresponding to five states due to
     degeneracies of energies $E_{n_2}$ and $E_{n_3}$.}
     \label{fig:sketchWKB}
\end{center}
\end{figure}

For an axially symmetric constriction aligned along the $z$-axis,
as depicted in Fig.\ \ref{fig:sketchWKB}, its geometry is
characterized by the $z$-dependent radius $R(z)$. Outside the
constriction, the solutions of the Schr\"odinger equation
decompose into plane waves along the wire and discrete eigenmodes
of a circular billiard in transverse direction.
The eigenenergies $E_{\mu\nu}$ of a circular billiard are given by
\begin{equation}
\label{e.nu}
    E_{\mu\nu}=\frac{\hbar^2}{2m_e}\frac{\gamma_{\mu\nu}^2}{\Rnull^2}\;,
\end{equation}
where the quantum number $\gamma_{\mu\nu}$ is the $\nu$-th root
of the Bessel function $\Jmu$ of order $\mu$ 
and $\Rnull$ is the radius of the wire outside the constriction.
In cylindrical coordinates $r$, $\varphi$, and $z$, the asymptotic
scattering states read
\begin{equation}
\label{inoutstates}
    \Psi_{\mu\nu}(r,\varphi,z) \sim e^{\pm{i}k_{\mu\nu}z+i\mu\varphi}\Jmu(\gamma_{\mu\nu} r/\Rnull)\;,
\end{equation}
where $k_{\mu\nu}(E)=\sqrt{{2m_e}(E-E_{\mu\nu})/{\hbar^2}}$ is the
longitudinal wavevector. In the following, we use multi-indices
$n=(\mu\nu)$ in order to simplify the notation.

If the constriction is smooth, i.e. $|\partial R/\partial z| \ll
1$, one may use an adiabatic approximation. In the adiabatic
limit, the transverse motion is separable from the motion
parallel to the $z$-axis even in the region of the constriction,
and the channel index $n$ of an incoming electron is preserved
throughout the wire. Accordingly, Eqs.\ (\ref{e.nu}) and
(\ref{inoutstates}) remain valid in the region of the constriction, with
$\Rnull$ replaced by $R(z)$.
The channel energies become functions of $z$,
$E_{n}(z)=\hbar^2\gamma_{n}^2/2m_eR(z)^2$, as is sketched in the
lower part of Fig.\ \ref{fig:sketchWKB}, and act as a potential
barrier for the effective one-dimensional scattering problem in
channel $n$. The corresponding Schr\"odinger equation for the
longitudinal part $\Phi$ of the wave function reads
\begin{eqnarray}
\label{eq:wkb:schroedingerC}
    \ddel{z}\,\Phi_n(z)+\frac{2m_e}{\hbar^2}\left[E-E_n(z)\right]\Phi_n(z)&=&0\;,
\end{eqnarray}
and is solved within the WKB approximation (see, e.g.,
\cite{Messiah-book}) by
\begin{eqnarray}
\label{eq:wkb:solutionC}
    \Phi_n(r,\varphi,z) \sim
    \frac{1}{\sqrt{k_n(E,z)}}
    \exp\left[\pm\,i\int_0^z k_n(E,z')dz'\right]\;.
\end{eqnarray}
For a constriction of length $L$ the transmission amplitude in
channel $n$ is then given by the familiar WKB barrier
transmission factor
\begin{eqnarray}\label{eq:tn}
    t_{n}(E)
    &=&
    \exp\left[{i\int_0^L dz\, k_n(E,z)}\right]
    \;\equiv\;
    \sqrt{\cT_n(E)}\,e^{i \Theta_n(E)}\,.
\end{eqnarray}
Here $\cT_n$ is the transmission coefficient of channel $n$ and
$\Theta_n$ is the corresponding phase shift. The transmission
amplitude gets exponentially damped in regions where the
transverse energy is larger than the state total
energy.\footnote{This simplest WKB treatment does not correctly
describe above-barrier reflection; a better approximation
including this effect is described by Brandbyge \ea (1995) and by
Glazman \ea (1988).}

The full S-matrix is now found to be of the form
\begin{equation}
    \label{eq:wkb:S-matrix}
    S=\left(\begin{array}{cc}
        i \sqrt{1-\cT}\,e^{i \Theta} & \sqrt{\cT}\,e^{i\Theta} \\
        \sqrt{\cT}\,e^{i \Theta} & i \sqrt{1-\cT}\,e^{i \Theta}
    \end{array}\right)\;,
\end{equation}
where for simplicity of notation we have suppressed the channel
indices and each of the entries is understood to be a diagonal
matrix in the channels. Using the formulas of Sec.
\ref{sec:SMatrix}, we may proceed to determine physical
quantities. From Eq. (\ref{condformula}) we deduce that the
electrical conductance at zero temperature reads,
\begin{eqnarray}
\label{eq:ConductanceWKB}
    G &=& \frac{2e^2}{h} \sum_n \cT_n(E_F)\\
      &=& \frac{2e^2}{h} \sum_n \exp\left[-2\int_0^L
    dz\,\theta(E_n(z)\!-\!E_F)\sqrt{\frac{2m_e}{\hbar^2}\left(E_n(z)-E_F\right)}\right],\nonumber
\end{eqnarray}
where the second line is obtained by using Eq.\ (\ref{eq:tn}).
Here $\theta(x)$ denotes the Heaviside step function
($\theta(x)=1$ for $x>0$, $0$ otherwise.) The density of states
is found to be connected with the phase shift $\Theta_n$,
\begin{eqnarray}\label{eq:DOS:WKB}
    D(E)&=&\frac{2}{\pi} \sum_n \Del{\Theta_n(E)}{E}
    \\
    &=&\frac{1}{\pi}\sqrt{\frac{2m_e}{\hbar^2}}\sum_n\int_0^L\!
    dz\,\frac{\theta(E-E_n(z))}{\sqrt{E-E_n(z)}}
    \,.
\end{eqnarray}

From the DOS, one gets the grand canonical potential in the limit
of zero temperature as
\begin{eqnarray}
\label{eq:OmegaWKB}
\Omega&&\stackrel{T\rightarrow0}{\longrightarrow}\;
   -\frac{8E_F}{3\la_F}
    \int_0^Ldz\,\sum_n\,\theta\big(E_F\!-\!E_n(z)\big)\left(1\!-\!\frac{E_n(z)}{E_F}\right)^{3/2}
\end{eqnarray}
which can then be used to calculate the tensile force and
stability of the nanowire, as discussed in the following sections.

\subsection{WKB approximation for non-axisymmetric wires}
The formalism presented in the previous subsection can be readily
extended to non-axisymmetric wires. In general, the surface of
the wire is given by the radius function $r=R(\varphi,z)$, which
may be decomposed into a multipole expansion
\begin{equation}
\label{eq:deformation}
    R(\varphi,z) = \rho(z)\left\{ \sqrt{1-\sum_m\frac{\la_m(z)^2}{2}}
      + \sum_m\la_m(z)\cos[m(\varphi\!-\!\varphi_m(z))]\right\},
\end{equation}
where the sums run over positive integers. The parameterization
is chosen in such a way that $\pi\rho(z)^2$ is the cross-sectional
area at position $z$. The parameter functions $\la_m(z)$ and
$\phi_m(z)$ compose a vector $\La(z)$, characterizing the
cross-sectional shape of the wire.

The transverse problem at fixed longitudinal position $z$ now
takes the form
\begin{eqnarray}
\label{eq:Seq:transverse}
    \left(\ddel{r}\! + \!\frac{1}{r}\del{r}
          \! + \!\frac{1}{r^2}\ddel{\varphi}\! + \!\frac{2m_e}{\hbar^2}E_n(z)\!\right)
      \chi_n(r,\varphi;z)&=&0,\quad
\end{eqnarray}
with boundary condition $\chi_n(R(\varphi,z),\varphi;z)=0$ for all
$\varphi\in[0,2\pi]$. This determines the transverse eigenenergies
$E_{n}(z)=E_{n}(\rho(z),\La(z))$ which now depend on the
cross-sectional shape through the boundary condition. With the
cross-section parametrization (\ref{eq:deformation}), their
dependence on geometry can be written as
\begin{eqnarray}
    E_{n}(\rho,\La)&=&\frac{\hbar^2}{2m_e}\left(\frac{\gamma_{n}(\La)}{\rho}\right)^2,
\end{eqnarray}
where the shape-dependent functions $\gamma_{n}(\La)$ remain to be
determined. In general, and in particular for non-integrable
cross-sections, this has to be done by solving Eq.\
(\ref{eq:Seq:transverse}) numerically \cite{Urban06}.

The adiabatic approximation (long-wavelength limit) implies the
decoupling of transverse and longitudinal motions. One starts with
the ansatz $\Psi(r,\varphi,z)=\chi(r,\varphi;z)\Phi(z)$ and
neglects all $z$-derivatives of the transverse wavefunction
$\chi$. Again one is left with a series of effective
one-dimensional scattering problems (Eq.\
\ref{eq:wkb:schroedingerC}) for the longitudinal wave functions
$\Phi_n(z)$, in which the transverse eigenenergies
$E_{n}\big(\rho(z),\La(z)\big)$ act as additional potentials for
the motion along the wire. These scattering problems can again be
solved using the WKB approximation and Eqs. (\ref{eq:DOS:WKB})
and (\ref{eq:OmegaWKB}) apply.

\subsection{Weyl-expansion}
\label{subsec:Weyl} Semiclassical approximations often give
an intuitive picture of the important physics and, due to their
simplicity, allow for a better understanding of some general
features. A very early analysis of the density of eigenmodes of a
cavity with reflecting walls goes back to Weyl (1911) who proposed
an expression in terms of the volume and surface area of the
cavity. His formula was later rigorously proved and further terms
in the expansion were calculated. Quite generally, we can express
any extensive thermodynamic quantity as the sum of such a
semiclassical Weyl expansion, which depends on geometrical
quantities such as the system volume $\cV$, surface area $\cS$,
and integrated mean curvature $\cC$, as well as an oscillatory
shell-correction due to quantum-size effects \cite{Brack97}. In
particular, the grand-canonical potential (\ref{eq:OmegaVonD})
can be written as
\begin{equation}
    \label{eq:OmegaWeyl}
    \Omega = -\omega\,\cV
    +\sigmaS\, {\cal S}
    -\gamma_s\,\cC
    +\delta\Omega,
\end{equation}
where the energy density $\omega$, surface tension coefficient
$\sigmaS$, and curvature energy $\gamma_s$ are, in general,
material- and temperature-dependent coefficients. On the other
hand, the shell correction $\delta\Omega$ can be shown, based on
very general arguments \cite{Strutinsky68,Zhang05}, to be a
single-particle effect, which is well described by the NFEM.

\subsection{Material dependence}
\label{sec:generalconstraint} Within the NFEM there is only one
parameter entering the calculation apart from the contact
geometry: the Fermi energy $E_F$, which is material dependent and
in general well known (see Tab.\ \ref{tab:sigma.gamma}).
Nevertheless, the energy cost of a deformation due to surface and
curvature energy, which can vary significantly for different
materials, plays a crucial role when determining the stability of
a nanowire.
Obviously, when working with a free-electron model, contributions
of correlation and exchange energy are not included, while they
are known to play an essential role in a correct treatment of the
surface energy \cite{Lang73}.
Using the NFEM {\it a priori} implies the macroscopic free energy
density $\omega=2E_Fk_F^3/15\pi^2$, the macroscopic surface energy
$\sigmaS=E_Fk_F^2/16\pi$, and the macroscopic curvature energy
$\gamma_s=2E_Fk_F/9\pi^2$. When drawing conclusions for metals
having surface tensions and curvature energies that are rather
different from these values, one has to think of an appropriate
way to include these material-specific properties in the
calculation.

A convenient way of modeling the material properties without
losing the pleasant features of the NFEM is via the
implementation of an appropriate surface boundary
condition. Any atom-conserving deformation of the structure is
subject to a constraint of the form
\begin{equation}
\label{eq:constraint}
  {\cal N} \equiv k_F^3{\cal V}
    - \eta_s\, 
      k_F^2{\cal S}
    + \eta_c\,k_F{\cal C}\, = 
       \rm{const}.
\end{equation}
This constraint on deformations of the nanowire interpolates
between incompressibility and electroneutrality as side
conditions, that is between volume conservation
($\eta_s=\eta_c=0$) and treating the semiclassical expectation
value for the charge $Q_\weyl$ \cite{Brack97} as an invariant
($\eta_s=3\pi/8$, $\eta_c=1$).

\begin{table}[]
\begin{center}
\begin{tabular}{lccccccc}
    Element                   &   Li  &  Na   &  K    &  Cu   &   Ag  &  Au   & Al  \\
\hline
    $E_F$ [eV]                &  4.74 &  3.24 &  2.12 &  7.00 &  5.49 &  5.53 & 11.7 \\
    $k_F$ [nm$^{-1}$]         &  11.2 &   9.2 &   7.5 &  13.6 &  12.0 &  12.1 & 17.5 \\
\hline
    $\sigmaS$ [meV/\AA$^{2}$] & 27.2  & 13.6  & 7.58  & 93.3  & 64.9  & 78.5   & 59.2   \\
    $\sigmaS$ [$E_Fk_F^2$]    & 0.0046& 0.0050& 0.0064& 0.0072& 0.0082& 0.0097 & 0.0017 \\
    $\eta_s$                    & 1.135 & 1.105 & 1.001 & 0.939 & 0.866 & 0.755  & 1.146  \\
\hline
    $\gamma_s$ [meV/\AA]      & 62.0  & 24.6  & 14.9  & 119   & 96.4  & 161    & 121    \\
    $\gamma_s$ [$E_Fk_F$]     & 0.0117& 0.0082& 0.0094& 0.0125& 0.0146& 0.0240 & 0.0059 \\
    $\eta_c$                     & 0.802 & 1.06  & 0.971 & 0.741 & 0.583 & -0.111 & 1.229  \\
\hline
\end{tabular}
\end{center}
\vspace*{-0.2cm} \caption[]{Material parameters
\onlinecite{Ashcroft-book,Perdew91} of several
  monovalent metals: Fermi energy $E_F$,
  Fermi wavevector $k_F$,
  surface tension $\sigmaS$,
  and curvature energy $\gamma_s$,
  along with the corresponding values of $\eta_s$ and $\eta_c$.
  The last column gives the corresponding values for the multivalent
  metal Al (see discussion in Sec.
  \ref{sec:materialdependence}.) Adapted from \cite{Urban06}.
 } \label{tab:sigma.gamma}
\end{table}

The grand canonical potential of a free-electron gas confined
within a given geometry by hard-wall boundaries, as given by Eq.\
(\ref{eq:OmegaWeyl}), changes under a deformation by
\begin{eqnarray}
    \Delta\Omega&=&
    -\omega\,\Delta\cV
    +\sigmaS\,\Delta\cS
    -\gamma_s\,\Delta\cC
    + \Delta[\delta \Omega]
    \\
    &=&-\frac{\omega}{k_F^3}\Delta\cN
    +\Big(\sigmaS-
    \frac{\omega}{k_F}\eta_s\Big)\Delta\cS
    -\Big(\gamma_s-\frac{\omega}{k_F^2}\eta_c\Big)\Delta\cC
    + \Delta[\delta \Omega],
    \nonumber
\end{eqnarray}
where the constraint (\ref{eq:constraint}) was used to eliminate
${\cal V}$. Now the prefactors of the change in surface
$\Delta\cS$ and the change in integrated mean curvature
$\Delta\cC$ can be identified as effective surface tension and
curvature energy, respectively. They can be adjusted to fit a
specific material's properties by an appropriate choice of the
parameters $\eta_s$ and $\eta_c$ (see Tab.\
\ref{tab:sigma.gamma}).

\section{Conductance and Force}\label{sec:ForceCond}

The formalism presented in the previous section can now be
applied to a specific wire geometry \cite{Stafford97a}, namely a
cosine constriction,
\begin{eqnarray}
\label{eq:cosineConstriction}
    R(z)&=&\frac{\Rnull\!+\!R_{min}}{2}\;+\;\frac{\Rnull\!-\!R_{min}}{2}\,\cos\left(\frac{2\pi z}{L}\right),
\end{eqnarray}
of a cylindrical wire. One is interested in the mechanical
properties of this metallic nanoconstriction in the regime of
conductance quantization. The necessary condition to have
well-defined conductance plateaus in a three-dimensional
constriction was shown \cite{Torres94} to be $(\partial R/\partial
z)^2 \ll 1$.  In this limit, one may employ the adiabatic and WKB
approximations and evaluate the expressions obtained in Sec.
\ref{sec:WKB}.

\begin{figure}[t]
    \begin{center}
       \includegraphics[width=0.8\columnwidth,draft=false]{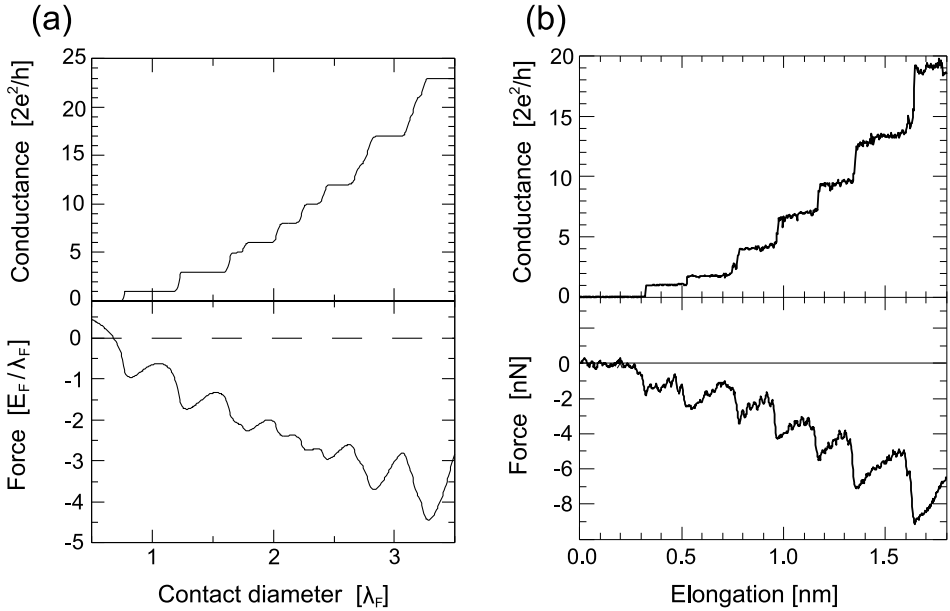}
    \end{center}
    \vspace{-0.5cm}
    \caption[]{\label{fig:GandF} Electrical conductance $G$ and
    tensile force $F$ as function of elongation of a nanowire,
    (a) calculated in a WKB approximation for a cosine
    constriction (adapted from Stafford \ea 1997), and (b) measured by Rubio \ea (1996) in an experiment on gold (data courtesy of N.\ Agra\"{\i}t).}
\end{figure}

\subsection{Conductance}
The conductance is obtained from Eq.\ (\ref{eq:ConductanceWKB}).
As the transmission amplitudes $\cT_n$ vary exponentially from 1
to 0 when the transverse energy of the respective channel at the
neck of the constriction traverses the Fermi energy, this results
in a steplike behavior of the conductance with almost flat
plateaus in between. This is the phenomenon of conductance
quantization, which is observable even at room temperature for
noble metal nanowires due to the large spacing of transverse
energies (of order 1 eV  for Au, to compare to
$k_BT\simeq10^{-3}$eV at room temperature). The upper panel of
Fig.\ \ref{fig:GandF}(a) shows the conductance obtained with an
improved variant of the WKB approximation
\cite{Brandbyge95,Glazman88} for the geometry
(\ref{eq:cosineConstriction}). The conductance as a function of
elongation shows the expected steplike structure and the step
heights are $2e^2/h$ and integer multiples thereof (the
multiplicity depends on the degeneracy of the transverse modes).
An ideal plastic deformation was assumed, i.e. the volume of the
constriction was held constant during elongation.\footnote{Note
the different abscissae for the theoretical and experimental
graphs in Fig.\ \ref{fig:GandF}. While both contact diameter and
elongation are a measure of the deformation of the contact, the
latter is more easily accessible to experiments. The contact
diameter on the other hand is the natural independent geometric
variable which is used for the theoretical graph, since its
relation to elongation is very model-dependent.}

\subsection{Force}
If the wire elongation is slow enough, the electron gas has time
to adjust to the wire shape changes during the deformation, and is
thus always in equilibrium.\footnote{The experimental elongation
speed is of order $1$nm/s, which is compatible with this
assumption.} Under these conditions, the tensile force can be
computed from the grand canonical potential, given in the WKB
approximation by Eq.\ (\ref{eq:OmegaWKB}), as
$F=-\partial\Omega/\partial L|_{\cal N}$.

The lower panel of Fig.\ \ref{fig:GandF}(a) shows the tensile
force for the cosine constriction~(\ref{eq:cosineConstriction}).
The correlations between the force and conductance are striking:
$|F|$ increases along the conductance plateaus, and decreases
sharply when the conductance drops. The constriction becomes
unstable when the last conductance channel is cut off, which is
indicated by a positive tensile force. Some transverse channels
are quite closely spaced, and in these cases, the individual
conductance plateaus [e.g., $G/G_0=14$, 15, 19, 21] and force
oscillations are difficult to resolve. The force oscillations are
found to have an amplitude $E_F/\lambda_F$ (i.e.\ $\sim 1.7$\ nN
for gold, consistent with experimental observations) independent
of the chosen geometry (circular and quadratic wires and cosine
and parabolic constrictions were tested), and to persist to very
large conductances.

These results can be understood within the intuitive picture of a
conductance channel as a delocalized metallic bond. The increase
of $|F|$ along the conductance plateaus and the rapid decrease at
the conductance steps can then be interpreted as stretching and
breaking of these bonds. Note that within the NFEM, the
correlation between conductance changes and force oscillations
comes from a pure quantum-size effect and not from atomic
rearrangements.

The comparison of theoretical predictions with experimental data
by Rubio \ea (1996), plotted in Fig.\ \ref{fig:GandF}(b), shows
very good qualitative agreement and underlines that, although the
NFEM is simple, it already gives a good qualitative description of
the experimental findings. Extensions of the NFEM including
structural dynamics of the wire (B\"urki \ea 2003, 2007a,b),
which are able to calculate the shape of the wire at all steps of
elongation, find that instabilities accelerate the conductance
and force jumps, making the theory even more similar to the
experiment.

In addition, it is easy to show that in the NFEM, the tensile
force is invariant under a stretching of the geometry $R(z)
\rightarrow R(\lambda z)$, so that
$F=({\varepsilon_F}/{\lambda_F})\;f(\Delta L/L_0, k_F R),$ where
$f(x,y)$ is a dimensionless function, i.e.\ the force oscillations
are universal, and thus do not depend on the details of the wire
shape, or precisely how it deforms. Nonuniversal corrections to
$F$ occur in very short constrictions, for which the adiabatic
approximation breaks down.

\section{Linear Stability Analysis}\label{sec:StabAna}

Metal nanowires are of great interest for nanotechnology, since
they may serve as conductors in future nanocircuits. In
particular, one would like to know whether a nanowire of given
length and radius remains stable at a given temperature.

At first sight, an atomistic approach seems to be more
``realistic'' than the NFEM and well suited to answer this
question. But molecular dynamics (MD) simulations conceptually are
not able to avoid the surface-tension-driven Rayleigh instability
of long nanowires. Since quantum-size effects from the electron
confinement are not properly taken into account, MD simulations
fail to give an explanation for the electronic shell and
supershell effects. On the other hand, atomistic quantum
calculations using, e.g., the local-density approximation, are
restricted to such small systems that their results can not really
be disentangled from finite-size effects \cite{Stafford00b}.
Therefore, to date, a stability analysis within the framework of
the NFEM (and generalizations thereof) is the only approach able
to correctly include the effects of electron-shell filling and
thereby shed light on the puzzling stability of long metal
nanowires.

The geometry of a wire of uniform cross section aligned along the
$z$-axis is characterized by the cross-sectional area ${\cal
A}=\pi\rho^2$, and a set of dimensionless parameters determining
the shape, which compose a vector $\La$ (cf.\ Eq.\
\ref{eq:deformation}). A small $z$-dependent perturbation of a
wire of length $L$ and initial cross section
$(\bar{\rho},\Labar)$ can be written in terms of a Fourier series
as
\begin{eqnarray}
  \label{eq:perturbation}
     \rho(z)  &=&\bar{\rho} + \eps\,\drho(z) \;=\; \bar{\rho} + \eps\sum_q\rho_q\,e^{i q z},
\nonumber\\
     \La(z) &=&\Labar + \eps\,\dLa(z)\;=\;\Labar + \eps\sum_q\La_q\,e^{i q z},
\end{eqnarray}
where the dimensionless small parameter $\eps$ sets the size of
the perturbation.\footnote{Assuming periodic boundary conditions,
the perturbation wave vectors $q$ must be integer multiples of
$2\pi/L$. In order to ensure that $\rho(z)$ and $\La(z)$ are
real, we have $\rho_{-q}=\rho^*_q$ and $\La_{-q}=\La^*_q$.}

The energetic cost of a small deformation of the wire can be
calculated by expanding the grand canonical potential as a series
in the parameter $\eps$,
\begin{equation}
\label{eq:omega.expand}
    \Omega=\Omega^{(0)}+\eps\,\Omega^{(1)}+\eps^2\,\Omega^{(2)}+{\cal O}(\eps^3).
\end{equation}
A nanowire with initial cross-section $(\bar{\rho},\Labar)$ is
energetically stable at temperature $T$ if and only if
$\Omega^{(1)}(\bar{\rho},\Labar,T)=0$ and
$\Omega^{(2)}(\bar{\rho},\Labar,T)>0$ for every possible
deformation $(\drho,\dLa)$ satisfying the constraint
(\ref{eq:constraint}).

\subsection{Rayleigh-Instability}
It is instructive to forget about quantum-size effects for a
moment and to perform a stability analysis in the classical limit.
For simplicity, one can restrict oneself to axial symmetry (i.e.
$\La\equiv0$). In the classical limit the grand canonical
potential is given by the leading order terms of the Weyl
approximation,
    $\Omega_\weyl=-\omega\,\cV+\sigmaS\, {\cal S}$,
and changes under the perturbation (\ref{eq:perturbation}) by
\begin{eqnarray}
\label{eq:deltaOmegaWeyl} \frac{\delta\Omega_\weyl}{L}
    &=&
    -2\pi \left(\bar{\rho}\,\omega-\sigmaS\right)\rho_0\eps
    +\pi \sum_{q\ne0}|\rho_q|^2
    \Big[-\omega+q^2\bar{\rho}\sigmaS\Big]\eps^2\;.
\end{eqnarray}
Because of the constraint (\ref{eq:constraint}) on possible
deformations, $\rho_0$ can be expressed in terms of the other
Fourier coefficients. Volume conservation, e.g., implies
$\rho_0=-({\eps}/{2\bar{\rho}})\sum_{q\ne0}|\rho_q|^2$ and
\begin{eqnarray}
\label{eq:deltaOmegaWeylcV} \frac{\delta\Omega_\weyl(q)}{L}
    &=&\frac{\pi\sigmaS}{\bar{\rho}}\sum_{q\ne0}|\rho_q|^2(\bar{\rho}^2q^2-1)\eps^2
\end{eqnarray}
which  has to be positive in order to ensure stability. Since $q$
is restricted to integer multiples of $2\pi/L$, stability requires
$L<2\pi\bar{\rho}$. This is just the criterion of the classical
Rayleigh instability \cite{Chandrasekhar81}: A wire longer than
its circumference is unstable and likely to break up into clusters
due to surface tension.

\subsection{Quantum-mechanical stability analysis}
\label{sec:QMStabAna} The crucial ingredient to the stabilization
of metal nanowires is the oscillatory shell correction
$\delta\Omega$ to the grand canonical potential
(\ref{eq:OmegaWeyl}) which is due to quantum-size effects. This
shell correction can be accounted for by a quantum-mechanical
stability analysis based on the WKB-approximation introduced in
Sec.\ \ref{sec:WKB}. The use of this approximation can be
justified  \emph{a posteriori} by a full quantum calculation
(Urban \ea 2003, 2007) which shows that the structural stability
of metal nanowires is indeed governed by their response to
long-wavelength perturbations. The response to short-wavelength
perturbations on the other hand controls a Peierls-type
instability characterized by the opening of a gap in the
electronic energy dispersion relation. This quantum mechanical
instability, which is missing in the semiclassical WKB
approximation, in fact limits the maximal length of stable
nanowires. Nevertheless, if the wires are short enough, and/or
the temperature is not too low, the full quantum calculation
essentially confirms the semiclassical results.

A systematic expansion of Eq.\ (\ref{eq:OmegaWKB}) yields
\begin{equation}
    \frac{\Omega^{(1)}}{L/\la_F}
    = 4\sum_{n} \sqrt{\frac{E_F-\bar{E}_{n}}{E_F}}
      \left(\La_0\!\cdot\!\bar{E}^{\prime}_n\,-\,2\bar{E}_{n}\frac{\rho_0}{\bar{\rho}}\right),
\label{eq:omega(1)}
\end{equation}
\begin{equation}
\label{eq:omega(2)}
  \frac{\Omega^{(2)}}{L/\la_F} = E_F
  \sum_q
    \binom{\rho_q/\bar{\rho}}{\La_q}^{\!\!\dagger} \!\!
    \left(\begin{array}{cc}
    A_{\rho\rho} & \!A_{\rho\La} \\ A_{\La\rho} &
    \!A_{\La\La}\end{array}\right)
    \!\!    \binom{\rho_q/\bar{\rho}}{\La_q},
\end{equation}
where the elements of the matrix $A$ in Eq.\ (\ref{eq:omega(2)})
are given by
\begin{eqnarray}
\label{eq:stabcoef:m}
  A_{\rho\rho}&=&\sum_{n} \frac{4\bar{E}_{n}}{E_F^{3/2}}\left[
    3\sqrt{E_F\!-\!\bar{E}_{n}}\,-\,\frac{\bar{E}_{n}}{\sqrt{E_F\!-\!\bar{E}_{n}}}
    \right],
\nonumber
\\
  A_{\La\rho}&=&-\sum_{n}\frac{4\bar{E}^{\prime}_n}{E_F^{3/2}}
    \left[
      \sqrt{E_F\!-\!\bar{E}_{n}}
      \,-\,\frac{\bar{E}_n}{2\sqrt{E_F\!-\!\bar{E}_n}}
    \right]\!,
\\
  A_{\La\La}&=&\sum_{n}\frac1{E_F^{3/2}}\left[
    2\bar{E}^{\prime\prime}_n\sqrt{E_F\!-\!\bar{E}_{n}}
    \,-\,\frac{\bar{E}^{\prime}_n\cdot(\bar{E}^{\prime}_n)^{\dagger}}
        {\sqrt{E_F\!-\!\bar{E}_{n}}}
    \right]\!.
\nonumber
\end{eqnarray}
Here $\bar{E}^{\prime}_n$ denotes the gradient of $E_{n}$ with
respect to $\La$ and $\bar{E}^{\prime\prime}_n$ is the matrix of
second derivatives. The bar indicates evaluation at
$(\bar{\rho},\Labar)$.

The number of independent Fourier coefficients in Eq.\
(\ref{eq:perturbation}) is restricted through the constraint
(\ref{eq:constraint}) on allowed deformations. Hence, after
evaluating the change of the geometric quantities $\cV$, $\cS$,
and $\cC$ due to the deformation, we can use Eq.\
(\ref{eq:constraint}) to express $\rho_0$ in terms of the other
Fourier coefficients, yielding an expansion
$\rho_0=\rho_0^{(0)}+\eps\,\rho_0^{(1)}+\cO(\eps^2)$. This
expansion then needs to be inserted in Eqs.\ (\ref{eq:omega(1)})
and (\ref{eq:omega(2)}), thereby modifying the first-order
change of the energy $\Omega^{(1)}$ and the stability matrix $A$
\cite{Urban06}.

Stability requires that the resulting modified stability matrix
$\tilde{A}$ be positive definite. Results at finite temperature
are obtained essentially in a similar fashion, by integrating
Eq.\ (\ref{eq:OmegaVonD}) numerically.

\subsection{Axial symmetry}\label{sec:AxialSymmetry}

\begin{figure}
    \begin{center}
    \includegraphics[width=0.8\columnwidth,draft=false]{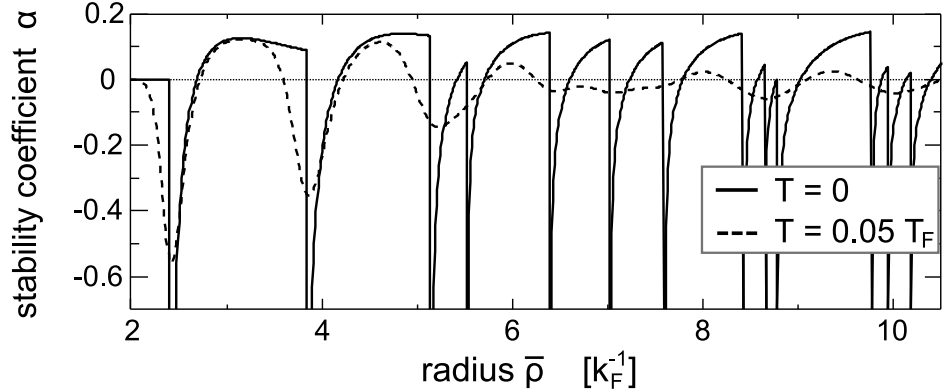}
    \end{center}
    \vspace{-0.5cm}
    \caption[]{\label{fig:alpha:WKB}
        WKB stability coefficient, calculated
        using a constant-volume constraint.
        The sharp negative peaks at the opening of new channels (i.e.\ when $k_F\bar{\rho}=\gamma_n$) are
        smeared out with increasing temperature $T$.}
\end{figure}

A straightforward application of the method outlined above is the
stability analysis of cylindrical wires with respect to
axisymmetric volume conserving perturbations. In this specific
case $\La(z)\equiv 0$ and
$\rho_0=-({\eps}/{2\bar{\rho}})\sum_{q\ne0}|\rho_q|^2$. Therefore,
Eq.\ (\ref{eq:omega(1)}) takes the form $\Omega^{(1)}\equiv0$ and
Eq.\ (\ref{eq:omega(2)}) simplifies to read
\begin{eqnarray}
    \frac{\Omega^{(2)}}{L/\la_F} &=& E_F
    \sum_{q\ne0}\left|{\rho_q}/{\bar{\rho}}\,\right|^2\alpha(\bar{\rho})\,,
\end{eqnarray}
where the \emph{stability coefficient}
$\alpha(\bar{\rho})\equiv\tilde{A}_{\rho\rho}$ reads
\cite{Urban06}
\begin{eqnarray}
    \alpha(\bar{\rho})&=&\sum_{n}\theta(k_F\bar{\rho}\!-\!\gamma_n)
    \frac{4\gamma_n^2}{(k_F\bar{\rho})^2}\left[4\sqrt{1\!-\!\frac{\gamma_n^2}{(k_F\bar{\rho})^2}}\,-\,\frac{1}{\sqrt{(k_F\bar{\rho})^2\!-\!{\gamma_n^2}}}\right].
\end{eqnarray}
Axial symmetry implies the use of the transverse eigenenergies
$E_n/E_F=(\gamma_n/k_F\bar{\rho}\,)^2$, cf.\ Eq.\ (\ref{e.nu}).
This result, valid for zero temperature, is plotted as a function
of radius in Fig.\ \ref{fig:alpha:WKB} together with a numerical
result at finite temperature. Sharp negative peaks at the subband
thresholds, i.e. when $\bar{\rho} k_F=\gamma_n$, indicate strong
instabilities whenever a new channel opens. On the other hand,
$\alpha$ is positive in the regions between these thresholds
giving rise to intervals of stability that decrease with
increasing temperature. These islands of stability can be
identified with the ``magic radii'' found in experiments. As will
be shown below, one has to go beyond axial symmetry in order to
give a full explanation of the observed conductance histograms of
metal nanowires.

\subsection{Breaking axial symmetry}\label{sec:JahnTeller}

It is well known in the physics of crystals and molecules that a
Jahn-Teller deformation breaking the symmetry of the system can be
energetically favorable. In metal clusters, Jahn-Teller
deformations are also very common, and most of the observed
structures show a broken spherical symmetry. By analogy, it is
natural to assume that for nanowires, too, a breaking of axial
symmetry can be energetically favorable, and lead to more stable
deformed geometries.

Canonical candidates for such stable non-axisymmetric wires are
wires with a $\cos(m\varphi)$-deformed cross section (i.e.\ having
$m$-fold symmetry), a special case of Eq.~(\ref{eq:deformation})
with only one non-zero $\la_m$. The quadrupolar deformation
($m=2$) is expected to be the energetically most favorable of the
multipole deformations\footnote{The dipole deformation ($m=1$)
corresponds, in leading order, to a simple translation, plus
higher-order multipole deformations. Therefore the analysis can be
restricted to $m>1$.} since deformations with $m>2$ become
increasingly costly with increasing $m$, their surface energy
scaling as $m^2$.

The results of Sec.\ \ref{sec:QMStabAna} can straightforwardly be
used to determine stable quadrupolar configurations by
intersection of the stationary curves,
$\Omega^{(1)}(\bar{\rho},\bar{\lambda}_2)|_{\cal N}=0$, and the
convex regions, $\Omega^{(2)}(\bar{\rho},\bar{\lambda}_2)|_{\cal
N}>0$. The result is a so-called \emph{stability diagram} which
shows the stable geometries (at a given temperature) in
configuration space, that is a as function of the geometric
parameters $\bar{\rho}$ and $\bar{\lambda}_2$. An example of such
a stability diagram is shown later in Fig.\ \ref{fig:AlStabDia}
for the case of aluminum, discussed in Sec.\
\ref{sec:materialdependence}. Results for all temperatures can
then be combined, thus adding a third axis (i.e.\ temperature) to
the stability diagram. Finally, the most stable configurations
can be extracted, defined as those geometries that persist up to
the highest temperature compared to their neighboring
configurations.

 \begin{table}[]
 \begin{center}
 \begin{tabular}{cccc}
       $G/G_0$ &
       $a$ &
       $\;\la_2$ &
       $T_{\rm{max}}/T_{\rho}$
 \\
 \hline
   2 & 1.72 & 0.26 & 0.50 \\
   5 & 1.33 & 0.14 & 0.49 \\
   9 & 1.22 & 0.10 & 0.50 \\
  29 & 1.13 & 0.06 & 0.54 \\
  59 & 1.11 & 0.05 & 0.49 \\
  72 & 1.08 & 0.04 & 0.39 \\
 117 & 1.06 & 0.03 & 0.55 \\
 172 & 1.06 & 0.03 & 0.50 \\
 \hline
 \end{tabular}
 \end{center}
    \caption{Most stable
    deformed wires with quadrupolar cross sections. The
    first column gives the quantized conductance of the corresponding
    wire. Both the aspect
    ratio $a$ and the value of the deformation parameter $\lambda_2$ are given. The
    maximum temperature of stability $T_{max}$ is given for each wire.
    In all cases the surface tension
    was set to $0.22\,\rm{N/m}$, corresponding to Na. Adapted from
    \cite{Urban06}.
    }\label{tab:DeformedWires}
 \end{table}

Table \ref{tab:DeformedWires} lists the most stable deformed
sodium wires with quadrupolar cross section, obtained by the
procedure described above. The deformation of the stable
structures is characterized by the parameter $\lambda_2$ or
equivalently by the aspect ratio
\begin{eqnarray}
    a&=&\frac{\sqrt{1-\lambda_2^2/2}\,+\,\lambda_2}{\sqrt{1-\lambda_2^2/2}\,-\,\lambda_2}.
\end{eqnarray}
Clearly, nanowires with highly-deformed cross sections are only
stable at small conductance. The maximum temperature up to which
the wires remain stable, given in the last column of Tab.\
\ref{tab:DeformedWires}, is expressed in units of
$T_\rho:=T_F/(k_F\bar{\rho})$. The use of this characteristic
temperature reflects the temperature dependence of the shell
correction to the wire energy \cite{Urban06}.

Deformations with higher $m$ cost more and more surface energy.
Compared to the quadrupolar wires, the number of stable
configurations with 3-, 4-, 5-, and 6-fold symmetry, their
maximum temperature of stability, and their size of the
deformations involved, all decrease rapidly with increasing order
$m$ of the deformation. For $m>6$ no stable geometries are known.
All this reflects the increase in surface energy with increasing
order $m$ of the deformation.

\subsection{General stability of cylinders}\label{sec:Cylinders}

It is possible to derive the complete stability diagram for
cylinders, i.e., to
determine the radii of cylindrical wires that are 
stable with respect to \emph{arbitrary} small, long-wavelength
deformations \cite{Urban06}. At first sight, considering arbitrary
deformations, and therefore theoretically an infinite number of
perturbation parameters seems a formidable task. Fortunately, the
stability matrix $\tilde{A}$ for cylinders is found to be
diagonal, and therefore the different Fourier contributions of the
deformation decouple. This simplifies the problem considerably,
since it allows to determine the stability of cylindrical wires
with respect to arbitrary deformations through the study of a set
of pure $m$-deformations, i.e.\ deformations as given by Eq.\
(\ref{eq:deformation}) with only one non-zero $\la_m$.

\begin{figure}[t]
\begin{center}
     \includegraphics[width=11.6cm,clip=,draft=false]{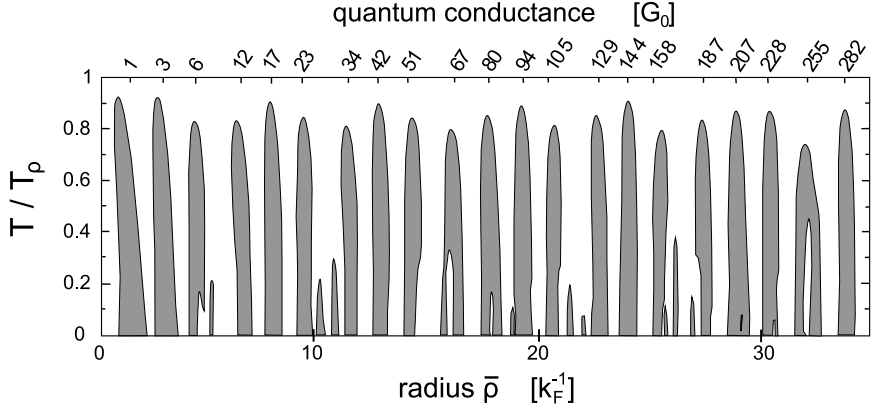}
\end{center}
      \caption[]{
      Stability of metal nanocylinders versus electrical conductance and
      temperature.  Dark gray areas indicate
      stability with respect to arbitrary small deformations.
      Temperature is displayed in units of $T_{\rho}=T_F/k_F\bar{\rho}$ (see text).
      The surface tension was taken as $0.22\,\rm{N/m}$, corresponding to
      Na. Adapted from \cite{Urban06}.
      } \label{fig:StabCylinders}
\end{figure}

Figure \ref{fig:StabCylinders} shows the stable cylindrical wires
in dark gray as a function of temperature. The surface tension was
fixed at the value for Na, see Tab.\ \ref{tab:sigma.gamma}. The
stability diagram was obtained by intersecting a set of
individual stability diagrams allowing
$\cos(m\varphi)$-deformations with $m\leq6$. This analysis
confirms the extraordinary stability of a set of wires with so
called ``magic radii''. They exhibit conductance values $G/G_0=$
1, 3, 6, 12, 17, 23, 34, 42, 51,... It is noteworthy that some
wires that are stable at low temperatures when considering only
axisymmetric perturbations, e.g., $G/G_0=$ 5, 10, 14, are found
to be unstable when allowing more general, symmetry-breaking
deformations.

The heights of the dominant stability peaks in Fig.\
\ref{fig:StabCylinders} exhibit a periodic modulation, with minima
occurring near $G/G_0=$ 9, 29, 59, 117, ... The positions of these
minima are in perfect agreement with the observed supershell
structure in conductance histograms of alkali metal nanowires
\cite{Yanson00}. Interestingly, the nodes of the supershell
structure, where the shell effect for a cylinder is suppressed,
are precisely where the most stable deformed nanowires are
predicted to occur (see discussion above). Thus symmetry breaking
distortions and the supershell effect are inextricably linked.

Linear stability is a necessary---but not a sufficient---condition
for a nanostructure to be observed experimentally.  The linearly
stable nanocylinders revealed in the above analysis are in fact
{\em metastable} structures, and an analysis of their lifetime has
been carried out within an axisymmetric stochastic field theory by
B\"urki \ea (2005a). There is a strong correlation between the
height of the stable fingers in the linear stability analysis and
the size of the activation barriers $\Delta E$, which determines
the nanowire lifetime $\tau$ through the Kramers formula
$\tau=\tau_0\exp(\Delta E/k_BT)$. This suggests that the linear
stability analysis, with temperature expressed in units of
$T_\rho=T_F/k_F\bar{\rho}$, provides a good measure of the total
stability of metal nanowires. In particular, the ``universal''
stability of the most stable cylinders is reproduced, wherein the
absolute stability of the magic cylinders is essentially
independent of radius (aside from the small supershell
oscillations).

\subsection{Comparison with experiments}\label{sec:compare}

\begin{figure}[]
  \begin{center}
       \includegraphics[width=0.82\columnwidth,draft=false,clip=]{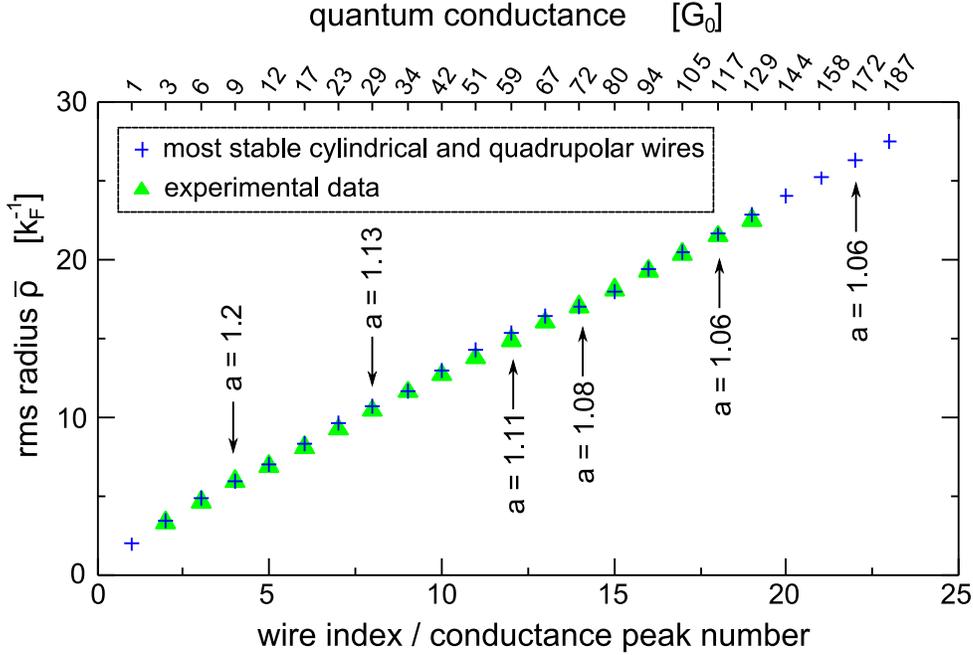}
  \end{center}
  \vspace{-0.5cm}
  \caption[]{\label{fig:compare}
    Comparison of the experimental shell structure
    for Na, taken from Yanson \ea (1999), with the theoretical
    predictions of the most stable Na nanowires.
    Non-axisymmetric wires are
    labeled with the corresponding aspect ratio $a$.
    Adapted from \cite{Urban04}.
    }
\end{figure}

A detailed comparison between the theoretically most stable
structures and experimental data for sodium is provided in Fig.\
\ref{fig:compare}. For each stable finger in the linear stability
analysis its mean conductance is extracted and plotted as a
function of its index number, together with experimental data by
Yanson \ea (1999). This comparison shows that there is a
one-to-one relation between observed conductance peaks and
theoretically stable geometries which in particular allows for a
prediction of the cross-sectional shape of the wires. This
striking fit is only possible when including non-axisymmetric
wires, which represent roughly 25\% of the most stable structures
and which are labeled by the corresponding aspect ratios $a$ in
Fig.~\ref{fig:compare}. The remaining $75\%$ of the principal
structures correspond to the magic cylinders. The role of symmetry
in the stability of metal nanowires is thus fundamentally
different from the case of atomic nuclei or metal clusters, where
the vast majority of stable structures have broken symmetry. The
crucial difference between the stability of metal nanowires and
metal clusters is not the shell effect, which is similar in both
cases, but rather the surface energy, which favors the sphere, but
abhors the cylinder.

Besides the geometries entering the comparison above, the
stability analysis also reveals two highly deformed quadrupolar
nanowires with conductance values of $2\;G_0$ and $5\;G_0$, cf.\
Tab.\ \ref{tab:DeformedWires}. They are expected to appear more
rarely due to
their reduced stability relative to the neighboring peaks, 
and their large aspect ratio $a$ that renders them rather
isolated in configuration space.\footnote{A nanowire produced by
pulling apart an axisymmetric contact has a smaller probability to
transform into a highly deformed configuration than into a
neighboring cylindrical configuration.} Nevertheless they can be
identified by a detailed analysis of conductance histograms of the
alkali metals \cite{Urban04b}.

\subsection{Material dependence}\label{sec:materialdependence}

Results for different metals are similar in respect to the number
of stable configurations and the conductance of the wires. On the
other hand, the deviations from axial symmetry and the relative
stability of Jahn-Teller deformed wires is sensitive to the
material-specific surface tension and Fermi temperature. The
relative stability of the highly deformed wires decreases with
increasing surface tension $\sigmaS/(E_Fk_F^2)$, measured in
intrinsic units, and this decrease becomes stronger with
increasing order $m$ of the deformation. Therefore, for the simple
s-orbital metals under consideration (Tab.\
\ref{tab:sigma.gamma}), deformed Li wires have the highest, and Au
wires have the lowest relative stability compared to cylinders of
``magic radii.''\footnote{Concerning the absolute stability, we
have to consider that the lifetime of a metastable nanowire also
depends on the surface tension \cite{Buerki05}.} Notable in this
respect is Aluminum with $\sigma_s=0.0017E_Fk_F^2$, some five
times smaller than the value for Au. Aluminum is a trivalent
metal, but the Fermi surface of bulk Al resembles a free-electron
Fermi sphere in the extended-zone scheme. This suggests the
applicability of the NFEM to Al nanowires, although the continuum
approximation is more severe than for monovalent metals.

\begin{figure}
  \begin{center}
    \includegraphics[width=0.7\textwidth]{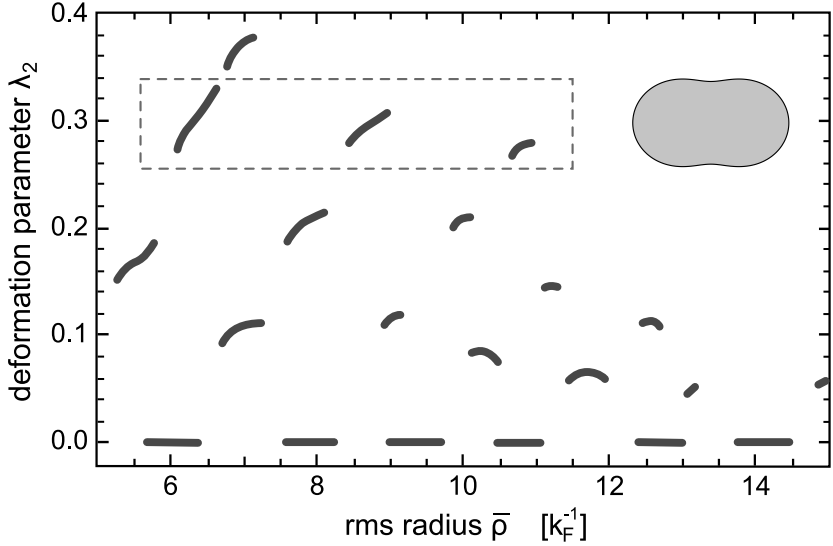}
  \end{center}
  \caption{Stability diagram for Al wires at fixed temperature $T=0.45\,T_\rho$.
    Thick lines mark stable wires in the configuration space of rms radius $\bar{\rho}$ and
    deformation parameter $\lambda_2$. The dashed box emphasizes a series of very stable
    superdeformed wires, whose peanut-shaped cross section is shown as an inset.
    This sequence was recently identified experimentally \cite{Mares07}.
  \label{fig:AlStabDia}}
\end{figure}

Recent experiments \cite{Mares07} have found evidence for the fact
that the stability of aluminum nanowires also is governed by shell
filling effects. Two magic series of stable structures have been
observed with a crossover at $G\simeq40G_0$ and the exceptionally

stable structures have been related to electronic and atomic shell
effects, respectively. Concerning the former, the NFEM can
quantitatively explain the conductance and geometry of the stable
structures for wires with $G>12G_0$ and there is a perfect
one-to-one correspondence of the predicted stable Al nanowires and
the experimental electron-shell structure. Moreover, an
experimentally observed third sequence of stable structures with
conductance $G/G_0 \simeq 5, 14, 22$ provides intriguing evidence
for the existence of ``superdeformed'' nanowires whose cross
sections have an aspect ratio near 2:1. Theoretically, these wires
are quite stable compared to other highly deformed structures and,
more importantly, are very isolated in configuration space, as
illustrated in the stability diagram shown in Fig.\
\ref{fig:AlStabDia}. This favors their experimental detection if
the initial structure of the nanocontact formed in the break
junction is rather planar with a large aspect ratio since then it
is likely that the aspect ratio is maintained as the wire necks
down elastically. Aluminum is unique in this respect and evidence
of superdeformation has not been reported in any of the previous
experiments on alkali and noble metals, presumably because
highly-deformed structures are intrinsically less stable than
nearly axisymmetric structures, due to their larger surface
energy.

\section{Summary and discussion}\label{sec:discussion}

In this chapter we have given an overview on the Nanoscale
Free-Electron Model, treating a metal nanowire as a
non-interacting electron gas confined to a given geometry by
hard-wall boundary conditions. At first sight, the NFEM seems to
be an overly simple model, but closer study reveals that it
contains very rich and complex features. Since its first
introduction in 1997, it has repeatedly shown that it captures the
important physics and is able to explain qualitatively, when not
quantitatively, many of the experimentally observed properties of
alkali and noble metal nanowires. Its strengths compared to other
approaches are, in particular, the absence of any free parameters
and the treatment of electrical and mechanical properties on an
equal footing. Moreover, the advantage of obtaining analytical
results allows the possibility to gain some detailed understanding
of the underlying mechanisms governing the stability and
structural dynamics of metal nanowires.

The NFEM correctly describes electronic quantum size effects,
which play an essential role in the stability of nanowires. A
linear stability analysis shows that the classical Rayleigh
instability of a long wire under surface tension can be completely
suppressed by electronic shell effects, leading to a sequence of
certain stable ``magic'' wire geometries. The derived sequence of
stable cylindrical and quadrupolar wires explains the
experimentally observed shell and supershell structures for the
alkali and noble metals as well as for aluminum. The most stable
wires with broken axial symmetry are found at the nodes of the
supershell structure, indicating that the Jahn-Teller distortions
and the supershell effect are inextricably linked. In addition, a
series of superdeformed aluminum nanowires with an aspect ratio
near 2:1 is found which has lately been identified experimentally.
A more elaborate fully quantum mechanical analysis within the NFEM
reveals an interplay between the Rayleigh and a Peierls-type
instability. The latter is length-dependent and limits the maximal
length of stable nanowires but other than that confirms the
results obtained by the long wavelength expansion discussed above.
Remarkably, certain gold nanowires are predicted to remain stable
even at room temperature up to a maximal length in the micrometer
range, sufficient for future nanotechnological applications.

The NFEM can be expanded by including the structural dynamics of
the wire in terms of a continuum model of the surface diffusion of
the ions. Furthermore, defects and structural fluctuations may
also be accounted for. These extensions improve the agreement with
experiments but do not alter the main conclusions. However, the
NFEM does not address the discrete atomic structure of metal
nanowires. With increasing thickness of the wire the effects of
surface tension decrease and there is a crossover from plastic
flow of ions to crystalline order, the latter implying atomic
shell effects observed for thicker nanowires. Therefore, the NFEM
applies to a window of conductance values between a few $G_0$ and
about $100G_0$, depending on the material under consideration.

Promising extensions of the NFEM in view of current research
activities are directed, e.g., towards the study of metal
nanowires in nanoelectromechanical systems (NEMS) which couple
nanoscale mechanical resonators to electronic devices of similar
dimensions. The NFEM is ideally suited for the investigation of
such systems since it naturally comprises electrical as well as
mechanical properties. It is hoped that the generic behaviour of
metal nanostructures elucidated by the NFEM can guide the
exploration of more elaborate, material-specific models, in the
same way that the free-electron model provides an important
theoretical reference point from which to understand the complex
properties of real bulk metals.

\section{References}

\enlarge

\end{document}